\documentclass[11pt,preprint]{emulateapj}
\usepackage{natbib}
\usepackage{color}
\usepackage{epsfig}
\usepackage{comment}
\usepackage{amsmath}
\usepackage{graphicx}
\shorttitle{Galaxies into the Dark Ages}

\shortauthors{Carilli, Murphy, Ferrara, Dayal}
 
\begin{document}

\title{Galaxies into the Dark Ages}

\author{
C. L. Carilli\altaffilmark{1,2},
 E.J. Murphy\altaffilmark{3},
A. Ferrara\altaffilmark{4,5},
P. Dayal\altaffilmark{6}
}

\altaffiltext{1}{National Radio Astronomy Observatory, P. O. Box 0,Socorro, NM 87
801, USA; ccarilli@nrao.edu}                                                     
\altaffiltext{2}{Astrophysics Group, Cavendish Laboratory, JJ Thomson Avenue, Cam
bridge CB3 0HE, UK}                                                              
\altaffiltext{3}{National Radio Astronomy Observatory, 520 Edgemont Road,        
Charlottesville, VA 22901, USA}                                                             
\altaffiltext{4}{Scuola Normale Superiore, Piazza dei Cavalieri 7, I-56126 Pisa, 
Italy}                                                                           
\altaffiltext{5}{Kavli IPMU, The University of Tokyo, 5-1-5 Kashiwanoha, Kashiwa
277-8583, Japan}
\altaffiltext{6}{Kapteyn Astronomical Institute, University of Groningen, P.O. Box 800, 9700 AV Groningen, The Netherlands}

\begin{abstract}

We consider the capabilities of current and future large facilities
operating at 2\,mm to 3\,mm wavelength to detect and image the [CII]
158\,$\mu$m line from galaxies into the cosmic `dark ages' ($z \sim
10$ to 20). The [CII] line may prove to be a powerful tool in
determining spectroscopic redshifts, and galaxy dynamics, for the
first galaxies. We emphasize that the nature, and even existence, of
such extreme redshift galaxies, remains at the frontier of open
questions in galaxy formation. In 40\,hr, ALMA has the sensitivity to
detect the integrated [CII] line emission from a moderate metallicity,
active star-forming galaxy [$Z_A = 0.2\,Z_{\odot}$; star formation
rate (SFR) = 5\,$M_\odot$\,yr$^{-1}$], at $z = 10$ at a significance
of 6$\sigma$. The next-generation Very Large Array (ngVLA) will detect
the integrated [CII] line emission from a Milky-Way like star
formation rate galaxy ($Z_{A} = 0.2\,Z_{\odot}$, SFR =
1\,$M_\odot$\,yr$^{-1}$), at $z = 15$ at a significance of
6$\sigma$. Imaging simulations show that the ngVLA can determine
rotation dynamics for active star-forming galaxies at $z \sim 15$, if
they exist. Based on our very limited knowledge of the extreme
redshift Universe, we calculate the count rate in blind, volumetric
surveys for [CII] emission at $z \sim 10$ to 20. The detection rates
in blind surveys will be slow (of order unity per 40\,hr
pointing). However, the observations are well suited to commensal
searches. We compare [CII] with the [OIII]
88$\mu$m line, and other ancillary information in high $z$ galaxies
that would aid these studies.

\end{abstract}


\section{Introduction}

The most sensitive observations with the largest telescopes at
$\gamma$-ray through radio wavelengths are now discovering galaxies,
AGN, and explosive phenomena in the redshift range $z \sim 6$ to 10,
some 940\,Myr to 500\,Myr after the Big Bang. This epoch corresponds
to `cosmic reionization', when light from early galaxies and accreting
black holes reionized the neutral intergalactic medium (IGM) that
pervaded the post-recombination Universe. Measurements of the cosmic
microwave background \citep{planck16}, the Gunn-Peterson
effect and related phenomena in the spectra of $z > 6$ quasars
\citep{banados16}, the Ly$\alpha$ emission line properties of $z
> 6$ galaxies \citep{ouchi17}, and most recently, limits to the HI
21cm emission from the neutral IGM at $z > 6$ \citep{parsons14,
ali15}, are narrowing the redshift range for cosmic reionization. It
is becoming clear that the $z \sim 6$ to 10 range corresponds to the
period during which the IGM transitions from mostly neutral, to highly
ionized, driven by early galaxy formation \citep{fan06, rob15,
greig17, dayal17}.
 
With the advent of the James Webb
Space Telescope (\textit{JWST}) and 30m-class ground-based optical and
near-IR telescopes, as well as implementation
of the full frequency range and capabilities of the Atacama Large
Millimeter Array (ALMA), we expect this important period of Universal
evolution to be well characterized over the coming decade.

What lies beyond? As we move toward the middle of the 21st century,
the redshift frontier will push back to $z \sim 10$ to 20,
corresponding to the epoch when the first stars and black
holes form, beginning the process of reionization, thereby
ending the cosmic Dark Ages \citep{lf13}.  

In this paper we explore the possibility of studying $z = 10$ to 20
galaxies using the 158\,$\mu$m fine structure line of ionized Carbon
with existing and future facilities operating at mm wavelengths.
Specifically, we consider the capabilities of the ever-improving ALMA,
as new receiver bands open the relevant redshift windows on the [CII]
line, in particular in the $z \sim 10-15$ range.  Pushing even further
out, to $z \sim 15-20$, we consider the capabilities of the `Next
Generation Very Large Array' (ngVLA) -- a facility being considered
for 2030 and beyond. The ngVLA takes the next order-of-magnitude leap
in sensitivity and resolution relative to the current cm and mm
facilities, required to study these first galaxies \citep{sm17, cc15, mcb16}.  

The $z \sim 15$ Universe is at the edge of our current understanding.
A handful of theoretical studies have speculated on the cosmic star
formation rate (SFR) density at these redshifts, in the context of
early reionization \citep{mashian16, duffy17, cp10, dayal14, yue15, 
topping15}.  The main difference
with lower redshift galaxy formation scenarios is probably related to
lower dynamical masses characterizing earlier structures.  This fact
makes them much more susceptible to supernova feedback which could
partially or totally suppress their star formation via gas ejection
and heating.  In addition, radiative feedback due to photo-ionizing
radiation emitted by nearby sources increases the Jeans length in the
intergalactic medium, therefore hampering the formation of the
smallest galaxies with circular velocities below $\approx 50$ km
s$^{-1}$ \citep{yue16, castellano16}.  
All these effects become increasingly important towards higher redshift.

Existing constraints on extreme redshift galaxies are poor, based on
extrapolation of the few galaxies and AGN known at $z \sim 7$ to 8,
and the even fewer galaxy candidates at $z \sim 8$ to 11.  An
encouraging observation is the indication of a relatively mature
interstellar medium and active star formation, in a few of the extreme
redshift sources discovered to date.  A very recent
result is the detection of the dust continuum and [OIII] 88$\mu$m fine
structure line emission from a candidate galaxy at $z = 8.4$
\citep{lapo17}. The galaxy is lensed modestly ($\mu \approx 2$),
with an intrinsic star formation rate of 20\,$M_\odot$\,yr$^{-1}$, a
stellar mass of $2 \times 10^9\,M_\odot$, and a dust mass of $6 \times
10^6\,M_\odot$. The star formation rate to stellar mass ratio places
this galaxy more than an order of magnitude above the standard `main
sequence' for star-forming disk galaxies in the nearby Universe. The
most extreme redshift candidate remains the $z \sim 11$ galaxy of
\citet{oesch16}.  If the source is at the stated redshift, the stellar
mass is $\sim 10^9\,M_\odot$, and the star formation rate is
24\,$M_\odot$\,yr$^{-1}$. While encouraging, observations remain
sparse and uncertain, and the most basic questions remain on the
nature, and even existence, of galaxies at $z \sim 15$.

Given the uncertainty in our knowledge of galaxies at extreme
redshifts, in this paper we focus on a few simple of questions: if
such extreme redshift galaxies exist, what kind of facility is
required to detect, and possibly image, the [CII] 158\,$\mu$m line
emission?  How do the prospects depend on basic galaxy properties,
such as metallicity and star formation rate? Based on what little we
know of galaxy demographics at very early epochs, what kind of
numbers can we expect in blind cosmological spectral deep fields?  We
do not consider lensing as a tool, but make the obvious point that
lensing can only help go to fainter galaxies
\citep[e.g.,][]{gullberg15}

The paper is organized as follows: In \S2 we describe the importance
of the [CII] 158\,$\mu$m fine structure and its promise as a way to
identify and characterize galaxies at $z \gtrsim 10$.  Then, in \S3 we
describe existing and future telescope capabilities to detect such
high-$z$ galaxies.  Our results and the implications are presented and
discussed in \S4.  Finally, our main conclusions are then summarized
in \S5.  All calculations are made assuming a Hubble constant $H_0 =
71\,\mathrm{km\,s}^{-1}\,\mathrm{Mpc}^{-1}$ and a flat $\Lambda$CDM
cosmology with $\Omega_{\rm M} = 0.27$ and $\Omega_{\Lambda} = 0.73$.

\section{Why the [CII] 158\,$\mu\rm m$ line?}
\label{sec:cii}

As we continue to push observations to more and more distant galaxies,
the standard rest-frame optical and UV spectral lines used
historically to determine redshifts move through the optical into the
near-IR windows. At $z > 10$, the Ly$\alpha$ line redshifts to an
observing wavelength $\lambda \ge 1.3\,\mu$m, and becomes increasing
difficult, or impossible, to observe from the ground. Moreover, even
from space, the Ly$\alpha$ line may be problematic due to the strong
resonant damping wings of Ly$\alpha$ absorption by the pervasive
neutral IGM at the end of the Dark Ages \citep{fan06}. Groups have
turned their attention to other atomic rest-frame UV-lines, such as
CIII] 1907, 1909 and [OII] 3726, 3729 \citep{barrow17}, to study the
first galaxies, in particular in the context of the up-coming \textit{JWST}.

In this paper, we consider millimeter observations, and using the
[CII] 158\,$\mu$m line.  The [CII] 158\,$\mu$m line is typically the
brightest of all spectral lines from star-forming galaxies at
far-infrared wavelengths and longer (although see \S\ref{sec:OIII}),
carrying between 0.1\% to 1\% of the total far infrared luminosity of star
forming galaxies \citep{stac91}. The [CII] fine structure line traces
both neutral and ionized gas in galaxies, and is the dominant coolant
of star-forming gas in galaxies \citep{pineda13, velusamy15,
langer14}.  Moreover, while the line is only visible from space in the
nearby Universe, it becomes easier to observe with increasing
redshift, moving into the most sensitive bands of large ground based
millimeter telescopes, such as NOEMA\footnote{\url
{http://iram-institute.org/EN/noema-project.php}}, and the
ALMA\footnote{\url{http://www.almaobservatory.org}}.

The last few years have seen an explosion in the number of [CII]
detections at high redshift, including high resolution imaging of the
gas dynamics on kpc-scales in distant galaxies.  The [CII] line is now
routinely detected in both AGN host galaxies and in more normal
star-forming galaxies at $z \sim 5.5$ to 7.5 \citep{willott15,
jones17, pentericci16, capak15, maiolino15, venemans16, 
cw13, bradac17, riechers13, riechers17, watson15, gullberg15,
strandet17, decarli16}.

Another important characteristic of the [CII] 158\,$\mu$m line is that
the ratio of [CII] luminosity to far-IR dust continuum luminosity
increases with decreasing metallicity \citep{pineda13}. The simple
point is that, once even a small amount of Carbon is present, it
becomes the dominant gas cooling line, hence balancing the heating by
star formation.

Considering emission line strength relative to the dust continuum
emission and the broad band sensitivity, the [CII] line-to-continuum
ratio (in terms of flux density), for $z \sim 6$ galaxies, has been
observed to be between 10 and 50 \citep{willott15, capak15, pentericci16}.  
The bandwidth for the line will be
limited to the line width, of order 100 km s$^{-1}$, or some 40\,MHz
at 110\,GHz observing frequency. Modern spectrometers are achieving
tens of GHz bandwidth, so the sensitivity is roughly 1000$^{1/2}$
better in the dust continuum, or a factor 30.  Hence, the detection
capabilities might be comparable for the line and continuum. However,
we focus on the [CII] line and not dust continuum for the following
reasons. First, the formation of dust within 500\,Myr of the Big Bang
remains highly uncertain, certainly not via mass loss from evolved AGB
stars \citep{michalowski10, dwek14, marassi15, schneider15}.  
Carbon is an $\alpha$ element, and hence
rapid ISM enrichment from the first generation of massive stars is
plausible on timescales $\le 100$ Myr. And second, the goal is not
just to detect the galaxy, but to determine its redshift, and possibly
the dynamics of the first galaxies.

\subsection{[CII] Luminosity, Metallicity, Star Formation, Redshift Relations}

As a predictor for the [CII] 158\,$\mu$m luminosity from early
galaxies we use the \citet{vallini15} relationship (their Equation
12). This theoretical and observational analysis considers in detail
the relationships between star formation rate, galaxy metallicity, and
[CII] luminosity to date. We adopt a few representative galaxy
characteristics, including the main parameters of: star formation
rate, metallicity, redshift, and [CII] luminosity, and compare these
to the capabilities of the given facilities.  We emphasize that the
detailed relationship between [CII] 158\,$\mu$m luminosity and star
formation rate is complex, and remains an area of active debate in the
literature, in particular at high redshift \citep{diaz-santos17,
  delooze14, olsen17}.

One of the chief unknowns is the metallicity of very early galaxies.
The obvious assumption would be low metallicity. However, there is
growing evidence for rapid build-up of metals in the early Universe,
at least in the denser regions of active structure formation.  Quasars
are seen with super-solar metallicity to $z \ge 6$
\citep{juarez09}.  Likewise, there are
galaxies, and galaxy candidates, with well developed ISM
characteristics, as seen through dust, CO, and atomic fine structure
line emission, at $z \sim 7$ to 8.4 \citep{venemans16, 
lapo17, riechers17, watson15}.  

In terms of current calculations of early galaxy formation, the
metallicity of the ISM of early galaxies is the ratio between the mass
of heavy elements produced by the stellar population and the hydrogen
mass, therefore being directly linked to the total mass of stars
formed over their assembly. As an example, we consider a galaxy of a
halo mass $M_h = 10^8 M_\odot$, corresponding to a $2 \sigma$
fluctuation at $z= 10$. Assuming a cosmological baryonic-to-dark
matter ratio equal to $\Omega_B / \Omega_M$, a conversion efficiency
of the gas into stars of $e*=0.03$ appropriate for early galaxies
\citep{dayal17}, and further taking the average metal yield per
stellar mass formed to be $y = 0.1$ M$_\odot$, we then get a gas
metallicity which depends only on the two last quantities, $Z = y e* = 
3 \times 10^{-3}$, or $Z \approx 0.15 Z_\odot$, 
essentially independent of halo mass or redshift. This
simple estimate assumes that metals and gas are perfectly mixed in the
galaxy and do not escape from it via outflows (i.e. a closed-box
solution), so it is perhaps an upper limit to $Z$ for a given
efficiency/yield. Such a situation is similar to what observed in the
BLR of high redshift quasars which always show a relatively high,
close to solar, metallicity even at very early times. This argument is
further supported by zoom-in numerical simulations
\citep{pallottini17}, or large scale ones \citep{wilkins17}. The
latter paper shows that for the smallest halos they can resolve
(stellar mass = $10^8$ M$_\odot$), going from $z=13$ to $z=8$, the gas
metallicity is bound in the range $-3.03 \le \log Z \le
-2.95$. Finally, as Carbon is rapidly produced in less than 100 Myr by
both pair-instability SNe and AGB stars, the time constraint is
relatively easy to achieve: a halo observed at $z=15$ must have formed
the first stars only by $z=20$.

Overall, the most likely scenario is that the very early Universe is
highly inhomogeneous on sub-Mpc-scales, with the densest regions
building up metals quickly, and lower density regions remaining
pristine \citep{wilkins17}.  In the present analysis we investigate a
similarly wide metallicity range to that used in \citet{vallini15},
i.e., $Z_{A} \sim 0.04, 0.2, {\rm and}~1.0\,Z_{\odot}$.

\section{Telescopes}

In the following section we consider the relevant capabilities for the
Atacama Large Millimeter Array, and the planned next generation Very
Large Array to detect the [CII] 158\,$\mu$m line at $z\gtrsim 10$ (see
Table \ref{tbl:facilities}).

\subsection{ALMA}
\label{sec:alma}

We assume that all the ALMA bands will be completed. In this case, the
relevant bands are 3, 4, and 5, corresponding to frequencies of
$84-116$\,GHz, $125-163$\,GHz, and $163-211$\,GHz, respectively. These
bands then cover the [CII] line (1900.54\,GHz rest frequency), between
$z = 10$ and 20, almost continuously. There is a gap due to
atmospheric O$_2$ absorption at $118\,$GHz with a width of a few MHz,
and a second strong atmospheric water line at 183\,GHz, with about
twice the width. The maximum frequency we consider is 173\,MHz. The
current bandwidth for ALMA is 8 GHz, although an increase to 32GHz is
being considered as a future development.

For ALMA sensitivity, we employ the ALMA sensitivity calculator, under
good weather conditions (3rd octile), with 50 antennas. For the sake
of illustration, we adopt a fiducial line width of 100 km s$^{-1}$
(see below), an on-source integration time of
40\,hr\footnote{Heretofore, all observing times quoted are on-source
onbserving time. Typical calibration overheads run at around 30\% to
40\%}, and a nominal observing frequency of 110\,GHz.  In this case,
the system temperature is $T_{\rm sys} \approx 75$\,K, and the rms
sensitivity per channel is 21 $\mu$Jy beam$^{-1}$ channel$^{-1}$.
Adopting the best weather (1st octile), only decreases $T_{\rm sys}$
to $\approx$73\,K.  The sensitivity of the array degrades with
increasing frequency, due to changing system temperature and system
efficiency. However, the line width also increases with frequency, in
terms of MHz for a fixed velocity width. These factors roughly offset
over the frequency range in question, implying comparable sensitivity
across the frequency range to within 10\%. For simplicity, we adopt
the value at 110\,GHz.  Lastly, we note that ALMA has multiple
configurations, all of which are designed to achieve a roughly
Gaussian synthesized beam shape for natural weighting of the
visibilities (= optimal sensitivity). We assume that the ALMA array
chosen is optimized for signal detection of the integrated emission
from the galaxies.

\begin{deluxetable}{ccccc}
\tablecaption{Facilities \label{tbl:facilities}}
\tablewidth{0pt}
\tablehead{
\colhead{Facilities} & Redshifts & \colhead{Frequencies} & \colhead{rms$^a$} & \colhead{Bandwidth}\\ 
\colhead{} &\colhead{} & \colhead{(GHz)} & \colhead{($\mu$Jy\,beam$^{-1}$)} & \colhead{(GHz)}
}
\startdata
ngVLA & $15 - 20$ & $116 - 90$ & 2.0 & 40 \\
ALMA & $10 - 15$ &  $173 - 116$ & 21 & 8 (32$^b$)
\enddata
\tablenotetext{a}{rms per channel in 40\,hr on-source 
and 100\,km\,s$^{-1}$ channel. } \\
\tablenotetext{b}{32GHz is a possible future upgrade to ALMA}
\end{deluxetable}

\subsection{A Next-Generation VLA}
\label{sec:ngvla}

The ngVLA is being considered as a future large radio facility
operating in the $\sim1.2-116$\,GHz
range\footnote{\url{https://science.nrao.edu/futures/ngvla}}.  The
current design involves ten times the effective collecting area of the
JVLA and ALMA, with ten times longer baselines ($\sim$300\,km)
providing milliarcsecond resolution, plus a dense core on a
1\,km-scale for high surface brightness imaging.  The ngVLA opens
unique new parameter space for imaging thermal emission from cosmic
objects ranging from protoplanetary disks to distant galaxies, as well
as unprecedented broad band continuum polarimetric imaging of
non-thermal processes \citep{mcb16, cc15}.

We employ the ``Southwest" configuration -- one of the proposed
configurations for the ngVLA \citep[][Greisen, Owen, \& Carilli {\sl
in prep}]{cc16}.  This array has 300 antennas distributed across New
Mexico, Chihuahua, and Texas.  The array includes 40\% of the antennas
in a core of diameter $\sim 1$\,km, centered on the VLA site. Then
some 30\% of the antennas out to VLA A-array baselines of 30\,km, and
the rest to baselines as long as 500\,km, into Northern Mexico and
Texas to enable AU-scale imaging of protoplanetary disks in nearby
star-forming regions.

For the ngVLA noise calculation, we adopt the interferometric
radiometer equation \citep{tms17}, 
using an 18\,m diameter antenna, with 70\% efficiency, 80\,K system temperature, a
40\,hr observation, and a 100 \,km s$^{-1}$ channel width.  We assume
observations from $90-115$\,GHz, implying a redshift range for [CII]
of $z \sim 15$ to 20.  The ngVLA bandwidth will cover this entire
range instantaneously.  Under these assumptions, we calculate a
naturally-weighted noise level of 1.3\,$\mu$Jy\,beam$^{-1}$\,channel$^{-1}$.

While the issue of reconfiguration of the ngVLA remains open, for this
exercise we conservatively assume a non-reconfigurable array.  The
current design of the ngVLA has a very non-uniform antenna
distribution.  The naturally-weighted beam for this centrally
condensed distribution leads to a PSF with a high resolution core of a
few mas width at 90\,GHz, plus a broad, prominent pedestal or plateau
in the synthesized beam with a response of $\sim 50\%$ over $\sim
1\arcsec$ scale. The goal in imaging is to adjust the relative
weighting of the data on different baselines lengths to obtain the
best the sensitivity, while maintaining a well behaved synthesized
beam (point spread function), relative to expected source sizes
(likely a few kpc, or $0\farcs1$ to $1''$). In our array simulations
below, we find that such a compromise can be reached on angular scales
relevant to the expected source sizes ($\sim 0\farcs2$ to $0\farcs4$),
with a loss of about a factor $\sim 1.5$ in sensitivity relative to natural
(optimal) weighting \citep{cc16}.  

\subsection{Simulations and Galaxy Parameters}
\label{sec:imgsim}

For the purpose of estimating the sensitivity of the ngVLA for
realistic observations, and to explore the imaging capabilities in the
event of the discovery of any relatively luminous sources, we have
employed the CASA simulation tools \citep{cc16, cc17}, developed for
the ngVLA
project\footnote{\url{https://science.nrao.edu/futures/ngvla/documents-publications}}.
We simulate a 40\,hr observation, made up of a series of 4\,hr
scheduling blocks around transit.

For imaging, we employ the CLEAN algorithm with Briggs weighting.  We
adjust the {\sc robust} parameter, the $(u,v)$-taper, and the cell
size, to give a reasonable synthesized beam and noise performance. Our
target resolution is $\sim 0\farcs4$ for detection, and $\sim 0\farcs2$ for
imaging.  The latter corresponds to 0.6\,kpc physical, at $z = 15$.

We adopt as a spatial and dynamical template, the observed CO 1-0
emission from the nearby star-forming disk galaxy, M\,51.  M\,51 is
one of the best studied galaxies in cool gas dynamics \citep{helfer03,
  schinnerer13} with a total observed line width of about 150 km
s$^{-1}$, and a disk radius in CO of about 5\,kpc. We assume that
rotational dynamics is the same for all gas constituents (e.g., CO or
[CII]). We also arbitrarily reduce the physical size of the disk by a
factor three, with the idea that very early galaxies are likely
smaller than nearby galaxies.  Again, this exercise is for
illustrative purposes, and the input model is just a representative
spatial/dynamical template for a disk galaxy, with the relevant
parameters being size, velocity, and luminosty. We employ the publicly
available BIMA SONG CO 1-0 data cubes \citep{helfer03}, as the
starting point of the models.  These data have high spatial resolution
(37\,pc) and excellent signal to noise.

We then adjust the line luminosity per channel per beam, to achieve a
given integrated [CII] 158\,$\mu$m luminosity at a given redshift.
The predicted luminosities as a function of basic galaxy properties
are discussed in the following section.

\section{Results}

In this section we present the results of our analysis to both detect
and characterize $z\gtrsim10$ galaxy candidates using ALMA and the
ngVLA, as well as searching for such high-$z$ sources via their [CII]
emission.

\subsection{Spectroscopic Confirmation of $z\gtrsim10$ Candidates}
\label{sec:specconf}

An obvious application of the [CII] 158\,$\mu$m line search will be to
determine spectroscopic redshifts for near-IR dropout candidate
galaxies at $z \sim 10$ to 20. Such spectroscopic verification using
[CII] may prove to be a powerful method to study of the earliest
galaxies, allowing for wideband searches and possible high resolution
imaging of the gas dynamics. The metallicity of
these galaxies remains an open issue, but on the positive side, the
galaxies most likely to be first discovered as near-IR dropouts by
\textit{JWST}, either in targeted deep fields or serendipitously, will
be the most prodigiously star-forming galaxies.  These will then be
the easiest to detect with the ngVLA and ALMA in their [CII] emission.

\begin{figure}[!t]
\centering 
\epsscale{1.25}
%\plotone{SCII.pdf}
\plotone{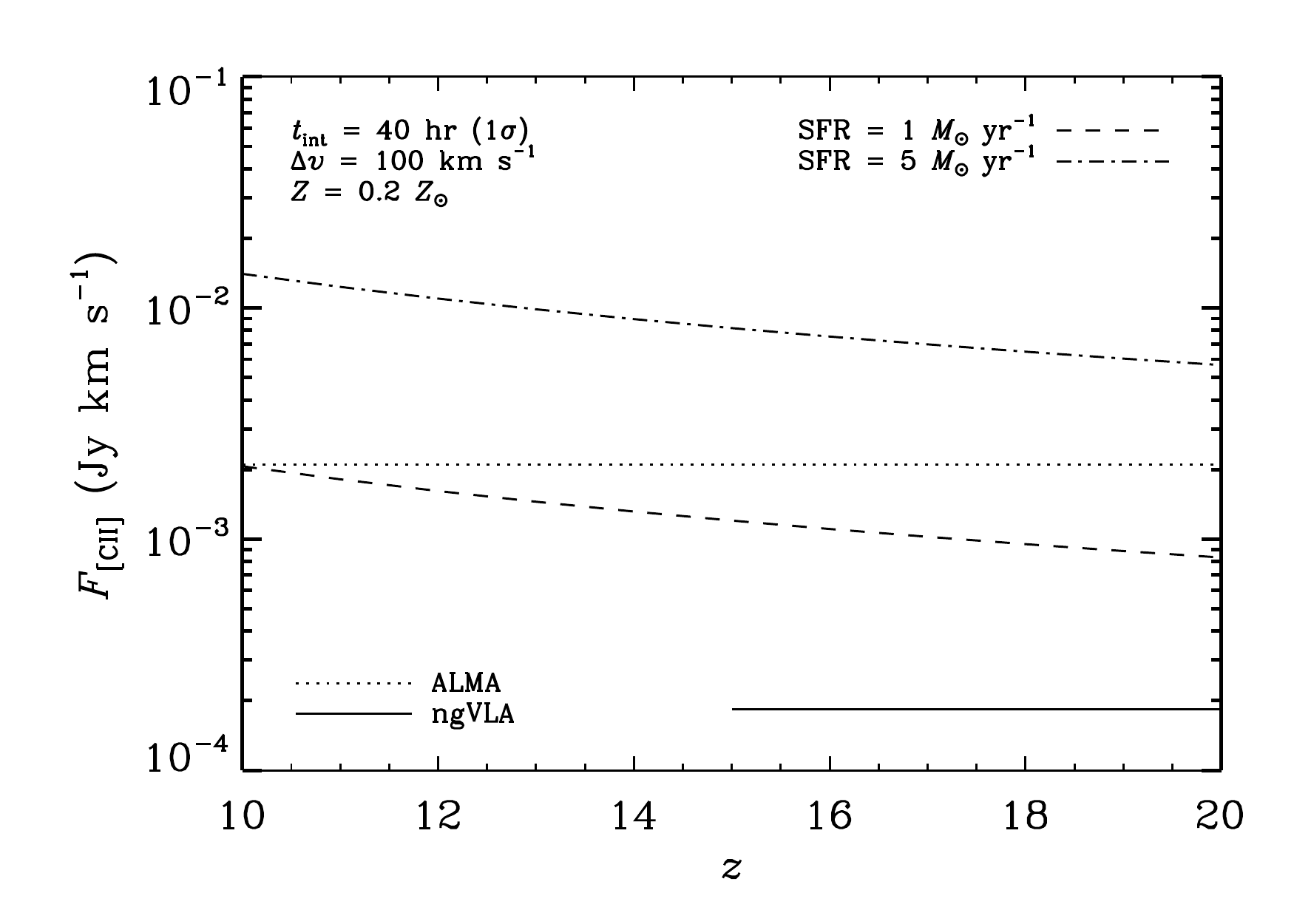}
\caption{[CII] 158\,$\mu$m velocity integrated line flux versus redshift
for galaxies with star formation rates of 1\,$M_\odot$\,yr$^{-1}$
and 5\,$M_\odot$\,yr$^{-1}$, and metallicity of 0.2\,$Z_{\odot}$,
 based on the relationship given in Equation 12 
of \citet{vallini15}. 
The rms sensitivity in a 100\,km\,s$^{-1}$ channel and 40\,hr integration
is shown for both ALMA and the ngVLA. 
}
\label{fig:scii}
\end{figure}

We start with the relationship between the [CII] velocity integrated
line flux, in the standard flux units of Jy km s$^{-1}$, versus
redshift. We adopt a metallically of $Z_A = 0.2\,Z_{\odot}$, and star
formation rates of 1\,$M_\odot$\,yr$^{-1}$ and
5\,$M_\odot$\,yr$^{-1}$. Figure~\ref{fig:scii} shows the predicted
[CII] line flux versus redshift for the two models, along with the
1$\sigma$ sensitivity of ALMA and the ngVLA. Again, we note that for
ALMA we adopt optimal (naturally weighted) sensitivity, assuming an
appropriate configuration is used for detection. For the ngVLA, we
have degraded the sensitivity by a factor 1.5 from optimal, due to
requirements of visibility weighting to obtain a reasonable PSF (see
\S \ref{sec:ngvla}).

\begin{figure}[!t]
\centering 
\epsscale{1.25}
%\plotone{LCII.pdf}
\plotone{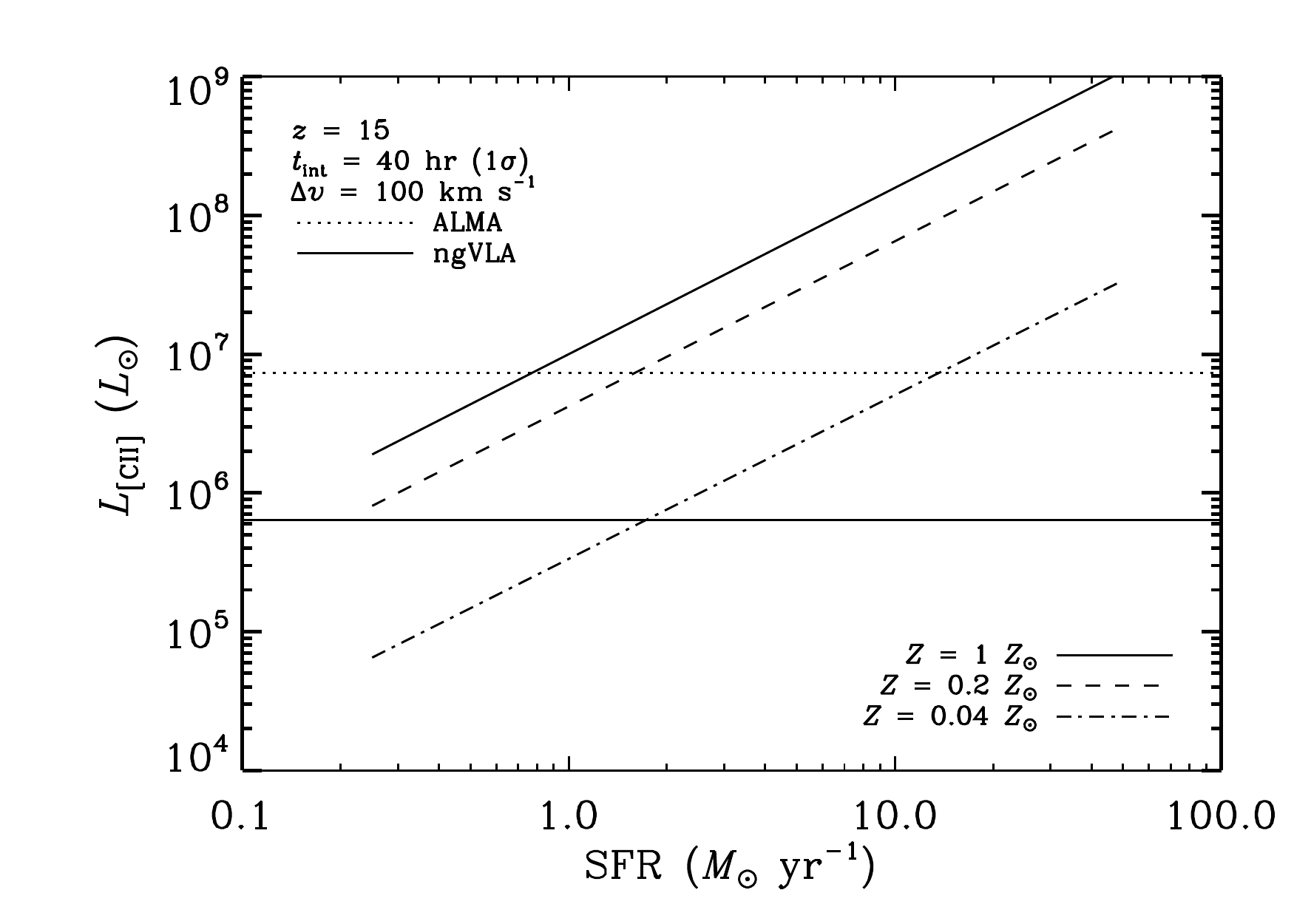}
\caption{[CII] 158\,$\mu$m line luminosity versus star formation 
rate and metallicity, based on the relationship given in Equation 12 
of \citet{vallini15}. Three different metallicities are shown.
Also shown is the rms sensitivity of ALMA and the ngVLA for
a galaxy at $z =  15$, assuming a 100\,km\,s$^{-1}$ 
channel and  40\,hr integration.
}
\label{fig:lcii}
\end{figure}

\begin{figure*}[!th]
\centering 
%\includegraphics[scale=0.35]{M5mod2.png}
%\plotone{M5mod2.png}
\plotone{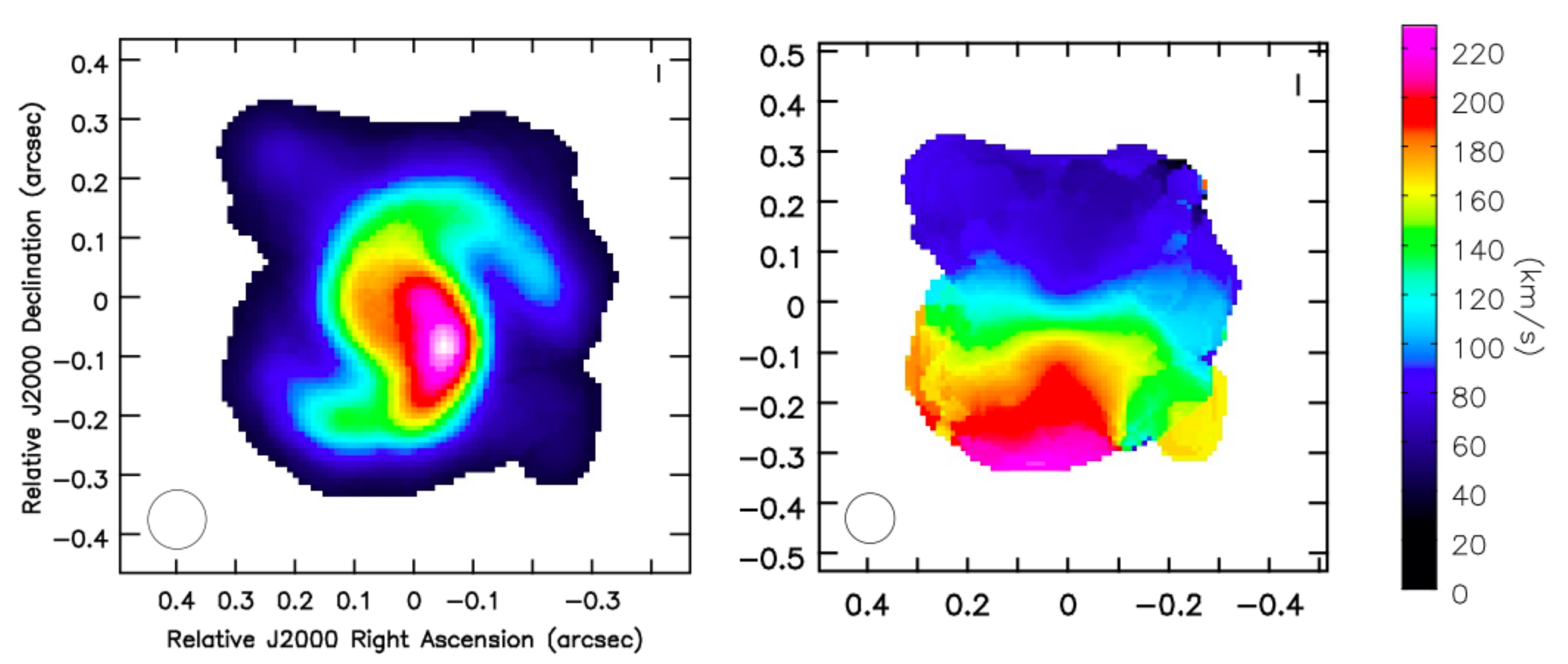}
\caption{
{\it Left:} A simulated image of the velocity integrated [CII]
158\,$\mu$m emission from a $z = 15$ galaxy with a star formation rate of
5\,$M_\odot$\,yr$^{-1}$, and a metallicity of 0.2\,$Z_{\odot}$. In this case,
no noise is added to the simulation, but the weighting applied 
to the visibilities was set to achieve a synthesized beam of 
FWHM $= 0\farcs1$, to obtain a better view of the intrinsic gas distribution
of the model. Left is the velocity integrated line emission. 
{\it Right:} The intensity weighted mean [CII] velocity (moment 1).
}
\label{fig:nonoise}
\end{figure*}
 
  \begin{figure}[!th]
\centering 
\plotone{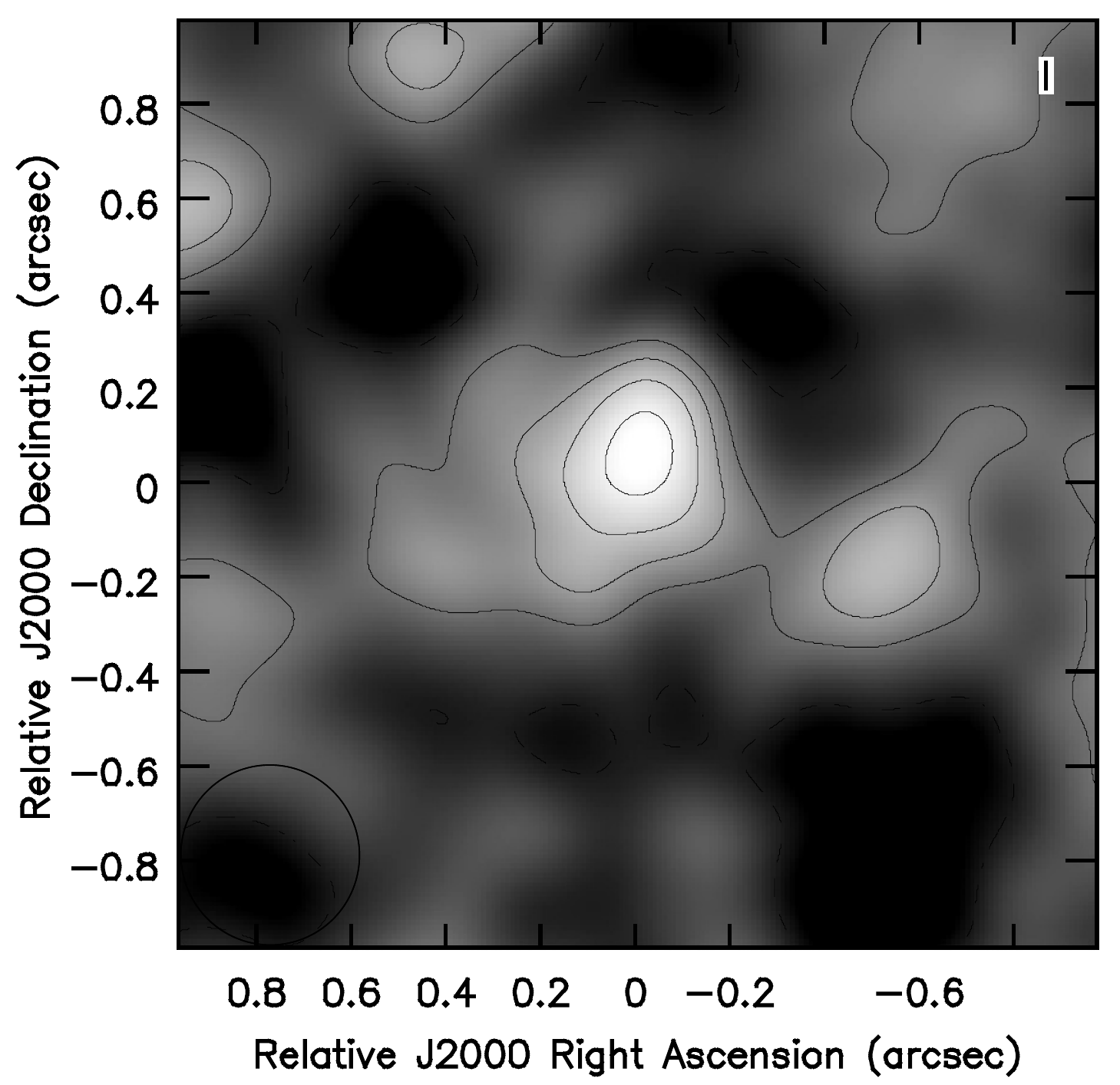}
\caption{
A simulated image of the velocity integrated [CII] 158\,$\mu$m
emission from a $z = 15$ galaxy with a star formation rate of
1\,$M_\odot$\,yr$^{-1}$, and a metallicity of 0.2\,$Z_{\odot}$,
assuming for a 40\,hr observation with the ngVLA. The contour levels
are -3.2,-1.6, 1.6, 3.2, 4.8, 6.4\,$\mu$Jy\,beam$^{-1}$. The rms noise on
the image is 1.6\,$\mu$Jy\,beam$^{-1}$, and the synthesized beam FWHM
is $0\farcs38$.
}
\label{fig:1M}
\end{figure}

In 40\,hr, the ngVLA will be able to detect the integrated [CII] line
emission from moderate metallicity and star formation rate galaxies
($Z_{A} = 0.2$, SFR = 1\,$M_\odot$\,yr$^{-1}$), at $z = 15$ at a
significance of 6$\sigma$. This significance reduces to 4$\sigma$ at
$z= 20$.

In 40\,hr, ALMA will be able to detect the integrated [CII] line
emission from a higher star formation rate galaxy ($Z_A =
0.2\,Z_{\odot}$, SFR = 5\,$M_\odot$\,yr$^{-1}$), at $z = 10$ at a
significance of 6$\sigma$. This significance reduces to 4$\sigma$ at
$z= 15$. ALMA will be hard-pressed to detect a moderate metallicity
($Z_A = 0.2\,Z_{\odot}$), lower star formation rate
(1\,$M_\odot$\,yr$^{-1}$) galaxy, requiring 1000\,hr for a 5$\sigma$
detection of the velocity integrated line flux, even at $z = 10$.

\begin{figure}[!th]
\centering 
\epsscale{1.2}
\plotone{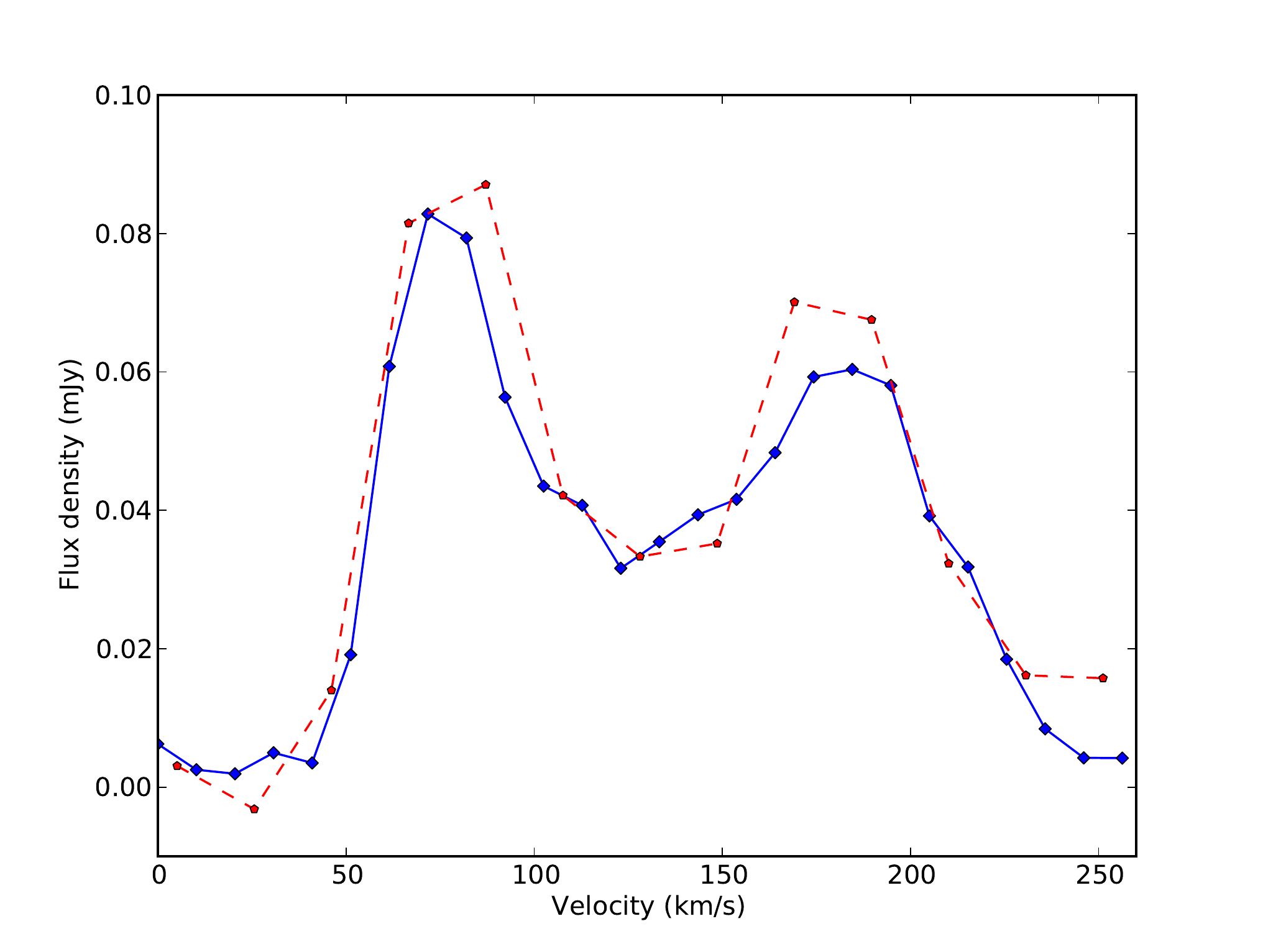}
\caption{
The red dashed line shows a simulated spectrum of the spatially
integrated [CII] 158\,$\mu$m emission from a $z = 15$ galaxy with a
star formation rate of 5\,$M_\odot$\,yr$^{-1}$, and a metallicity of
0.2\,$Z_{\odot}$, assuming for a 40\,hr observation with the ngVLA, at
20\,km\,s$^{-1}$\,channel$^{-1}$. The blue line shows the same
spectrum, but with no noise added and at
10\,km\,s$^{-1}$\,channel$^{-1}$.
}
\label{fig:5Mspec}
\end{figure}

\begin{figure*}[!th]
\centering 
%\includegraphics[scale=0.35]{5}
%\plotone{M5.png}
\plotone{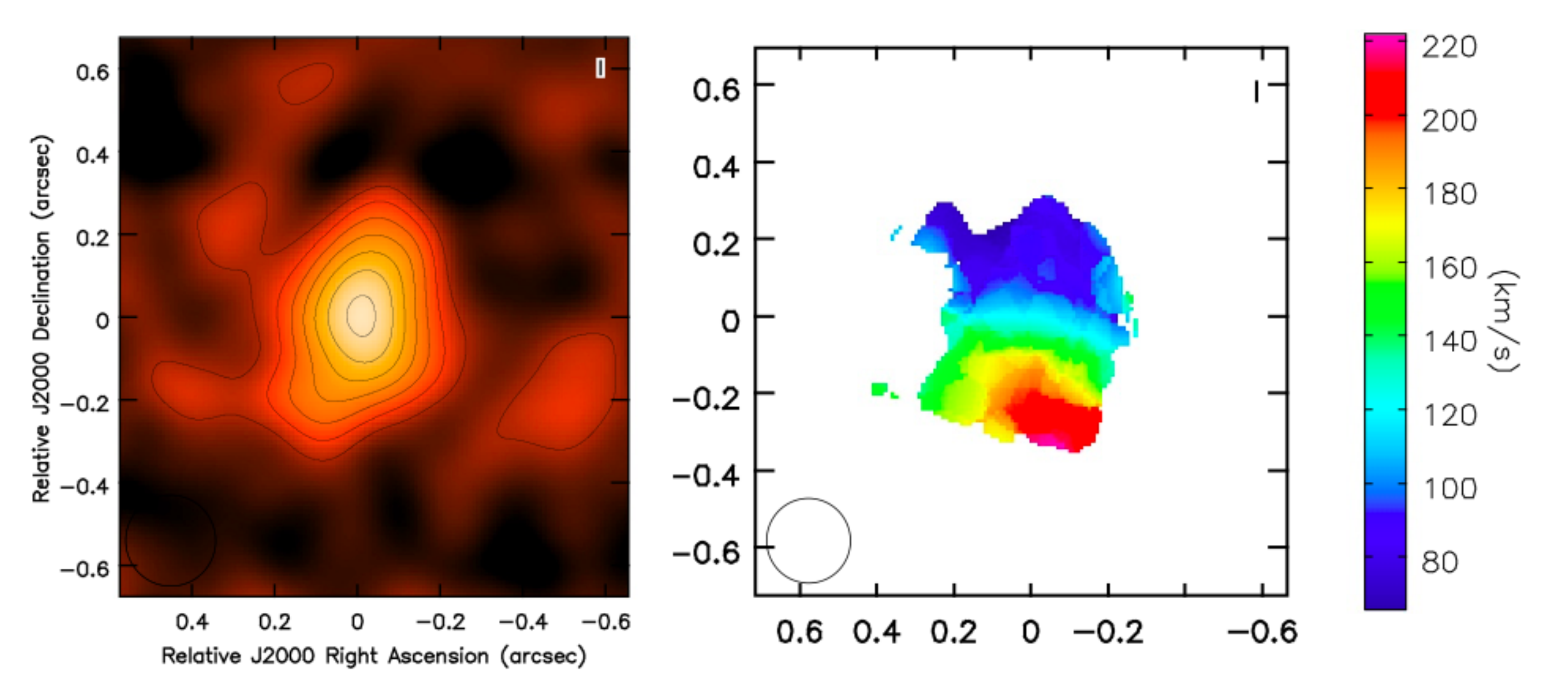}
\caption{
{\it Left:} A simulated image of the velocity integrated [CII]
158\,$\mu$m emission from a $z = 15$ galaxy with a star formation rate
of 5\,$M_\odot$\,yr$^{-1}$, and a metallicity of 0.2\,$Z_{\odot}$,
assuming for a 40\,hr observation with the ngVLA. Left is the velocity
integrated line emission.  The contour levels are
-6, -3, 3, 6, 9, 12, 15, 18, 21\,$\mu$Jy\,beam$^{-1}$. The rms noise on the
image is about 1.8\,$\mu$Jy\,beam$^{-1}$, and the synthesized beam
FWHM is $0\farcs22$.  {\it Right:} The intensity weighted mean [CII]
velocity (moment 1).
}
\label{fig:5M}
\end{figure*}

We next consider dependence on metallicity.  Figure~\ref{fig:lcii}
shows the relationship between [CII] luminosity (in Solar units), to
star formation rate, for three different metallicities: $Z_A = 0.04$,
0.2, and 1.0\,$Z_{\odot}$, for a galaxy at $z = 15$. Again shown
are the ALMA and ngVLA sensitivities in 40\,hr, 100 km s$^{-1}$
channels. The \citet{vallini15} model has the [CII] luminosity as a strong
function of metallicity.  If the gas has Solar metallicity, the ALMA
detection threshold ($4\sigma$) reduces to a galaxy with a star
formation rate of 2.5\,$M_\odot$\,yr$^{-1}$ (compared to
5\,$M_\odot$\,yr$^{-1}$ for $Z_A = 0.2$), while that for the ngVLA
reduces to 0.4\,$M_\odot$\,yr$^{-1}$ (compared to
1\,$M_\odot$\,yr$^{-1}$ for $Z_A = 0.2$). Conversely, for a low
metallicity galaxy of $Z_A = 0.04\,Z_{\odot}$, these values increase
to 100\,$M_\odot$\,yr$^{-1}$ and 10\,$M_\odot$\,yr$^{-1}$,
respectively.

Consequently, it appears that ALMA should be able to spectroscopically
confirm drop-out candidate galaxies forming stars at a rate of a few
solar masses per year with metallicity $\ge 0.2$ at $z \sim 10$, in
reasonable integration times.  The ngVLA pushes this detection limit
to $z \sim 15$ to 20, for star formation rates of order unity with
$Z_A \ge 0.2$.  If such galaxies do exist, ALMA and the
ngVLA should be excellent tools to confirm their existence.

\subsection{Kinematics of $z\gtrsim10$ Galaxies }

We investigate the potential for obtaining kinematic information
from such galaxies using the ngVLA.  We start by considering
visibility weighting to obtain a detection of the integrated emission
from a high redshift galaxy with the configuration of the ngVLA. The
imaging is a complex optimization procedure, balancing the Briggs {\sc
robust} weighting parameter \citep{briggs95}, 
the Gaussian tapering of the
($u,v$)-weighting, and the cell size in the gridding kernel, to
approach a reasonable balance between good sensitivity and the
behaviour of the PSF. Pure natural weighting for
the ngVLA leads to a PSF `core' of just a few milliarcseconds due to
the 300\,km baselines, which radically over-resolves the emission.
See Carilli (2016) for more details on imaging optimization using the
suite of current tools in CASA.

We have explored a few of the main parameters using the tools
available, with a goal of getting a rough estimate of the loss of
sensitivity when imaging with non-optimal array configurations.  We
expect the search for optimal imaging techniques for various goals
(simple detection or high resolution imaging), to be a long-term
exercise in interferometric imaging, with the advent of the complex
array configurations envisioned for facilities such as the ngVLA and
the Square Kilometer Array. Our current estimates of
sensitivity are likely conservative, depending on future algorithmic
development.

For reference, Figure~\ref{fig:nonoise} shows results for the input
galaxy model we use to explore the imaging parameters, as discussed in
\S\ref{sec:imgsim}. In this case, we have imaged the source with
no noise added, and using imaging parameters that result in a PSF with
a FWHM = $0\farcs1$, in order to show the intrinsic properties of the
model galaxy. We show both the velocity integrated [CII] emission, and
the intensity weighted mean [CII] velocity.  The model shows spiral
arms extending over an area of about $\sim 0\farcs4$, with the
majority of the emission centrally condensed bar and nucleus in the
inner $\sim 0\farcs2$.

Figure~\ref{fig:1M} shows the image of the velocity integrated [CII]
line emission from the $Z_A = 0.2\,Z_{\odot}$, and SFR =
1\,$M_\odot$\,yr$^{-1}$ galaxy, assuming noise appropriate for a
40\,hr observation. We adopt imaging parameters that optimize
detection of the integrated emission.  The emission is clearly
detected using Briggs weighting with {\sc robust} = 1, a Gaussian
($u,v$)-taper of $0\farcs2$, and a cell size of $0\farcs01$. This yields a
beam FWHM $\sim 0\farcs4$ and an rms of $1.6\,\mu$Jy beam$^{-1}$ over
the 150\,km\,s$^{-1}$ velocity range (or about $2\,\mu$Jy beam$^{-1}$
at 100 km s$^{-1}$ channel$^{-1}$, compared to $1.3\,\mu$Jy
beam$^{-1}$ for natural weighting of the visibilities). The result is
about a $5.5\sigma$ detection of the integrated emission from Gaussian
fitting.

We next consider imaging of the higher star formation rate model, with
$Z_A = 0.2\,Z_{\odot}$, and 5\,$M_\odot$\,yr$^{-1}$ galaxy for a
40\,hr observation. Given the brighter signal, we investigate whether
information on the gas dynamics can be recovered with high resolution
imaging.  We employ Briggs weighting with {\sc robust} = 0.5, a
Gaussian ($u,v$)-taper of $0\farcs15$, and a cell size of $0\farcs01$. This
yields a beam FWHM $\sim 0\farcs2$.  We synthesize channel images at
20\,km s$^{-1}$\,channel$^{-1}$, for which the rms noise is about
4.5\,$\mu$Jy beam$^{-1}$.  We also generate a velocity integrated
[CII] image averaging over the full width of the line.

The resulting spectrum, integrated over the source area, is shown in
Figure~\ref{fig:5Mspec}.  The red dash line is the simulated spectrum
at 20\,km\,s$^{-1}$\,channel$^{-1}$ with noise added, while the blue line
shows the integrated line emission made from data with no noise added,
and at 10\,km\,s$^{-1}$\,channel$^{-1}$, as a reference spectrum (Fig
3). Clearly, the ngVLA can make a high signal to noise detection of the
emission from this galaxy, with an integrated significance for the
detection of about 20$\sigma$.

From the channel images we generate the intensity weighted mean
velocity image (moment 1), using surface brightnesses above
2$\sigma$. The result is shown Figure~\ref{fig:5M}. The velocity
integrated intensity, and mean velocity, images can be compared to
Figure 4, which again shows the same model, but with noiseless ($u,v$)
data, and at higher spatial resolution. Clearly, at this signal to
noise and resolution we cannot recover the detailed structure of the
gas, such as the spiral arm features. However, the overall velocity
gradient is recovered, including the maximum and minimum velocity of
the gas, as well as the north-south orientation and extension of the
major axis.

\subsection{The Potential for Blind Searches of $z\gtrsim10$ Galaxies}

Another application for the [CII] line will be blind cosmological deep
fields. The advent of very wide bandwidth spectrometers has led to a
new type of cosmological deep field, namely, spectral volumetric deep
fields, in which a three dimensional search for spectral lines can be
made, with redshift as the third dimension (e.g., Walter et al. 2016).

To this aim we consider two predictions for the number density of
galaxies at these very high redshifts from the recent literature.
These predictions employ very different methodologies. Again, we point
out that the current observational constraints are extremely limited.
Both models employ a Salpeter IMF from 0.1 to 100 M$_\odot$.

First, we consider the galaxy number counts of \citet[][CP10]{cp10}.  
These galaxy counts are based on backward-evolving models for
the infrared luminosity function of \citet{ce01}, anchored by a
variety of observational data including the deepest {\it Spitzer}
24\,$\mu$m imaging from the GOODS fields, the fraction of the
far-infrared background light resolved by {\it Spitzer} and {\it
Herschel}, spectroscopic redshifts of {\it Spitzer} and {\it Herschel}
sources in the deep fields, and are consistent with the number counts
as well as $P(D)$ analysis from deep {\it Herschel} observations.

Second, we employ the calculation of high redshift galaxy formation of
\citet[][Dayal14]{dayal14}.  This model aims at isolating the
essential physics driving early galaxy formation via a merger-tree
based semi-analytical model including the key physics of star
formation, supernova feedback and the resulting gas ejection, and the
growth of progressively more massive systems (via halo mergers and gas
accretion). It involves only two free parameters, the star formation
efficiency threshold, $f_*$, and the fraction of SN energy that drives
winds, $f_w$. The key premise is that any galaxy can form stars with a
maximal effective efficiency, $f_*^{eff}$, that provides enough energy
to expel all the remaining gas, quenching further star formation. The
value of $f_*^{eff} = min[f_*,f_*^{ej}]$ where $f_*^{ej}$ is the
star-formation efficiency required to eject all gas from a
galaxy. Thus, low-mass galaxies form stars at a more limited
efficiency than massive galaxies.

The model has been validated against available high-$z$
data. For example, it reproduces extremely well both the slope and
amplitude of the UV LF from $z=5$ to $z=10$ at the same time providing
a physical explanation for the slope evolution in terms of a faster
assembly of galaxies at earlier redshifts.  Dayal14 also predicts that
the bright-end slope of the UV LF should be flatter than the steep
drop-off implied by the Schechter function, and actually closer to
the slope of the underlying dark matter halo mass function. This, in
turn, might be interpreted as a limited impact of quasar feedback at
high redshifts.

The two models predict the cummulative co-moving number density of
star-forming galaxies above a given star formation rate as a function
of redshift. We show the results in Figures 7a and 7b for the CP10 and
Dayal14 models, respectively. 

The co-moving number densities can be turned into the number of
observed galaxies in a given integration time, bandwidth, and field of
view, using the sensitivities of the ngVLA and ALMA. In \S\ref{sec:specconf}, we calculated that, in 40\,hr for a galaxy with
$Z_{A} = 0.2$, the ngVLA can detect a SFR = 1\,$M_\odot$\,yr$^{-1}$
galaxy at 6$\sigma$ significance between $z = 15$, reducing to
4$\sigma$ at $z= 20$. ALMA can detect a SFR = 5\,$M_\odot$\,yr$^{-1}$
galaxy at 6$\sigma$ at $z = 10$, reducing to 4$\sigma$ significance
between $z = 15$.  We use these two star formation rates for
demonstrative purposes.

\begin{figure}[!t]
\centering 
\epsscale{1.2}
\plotone{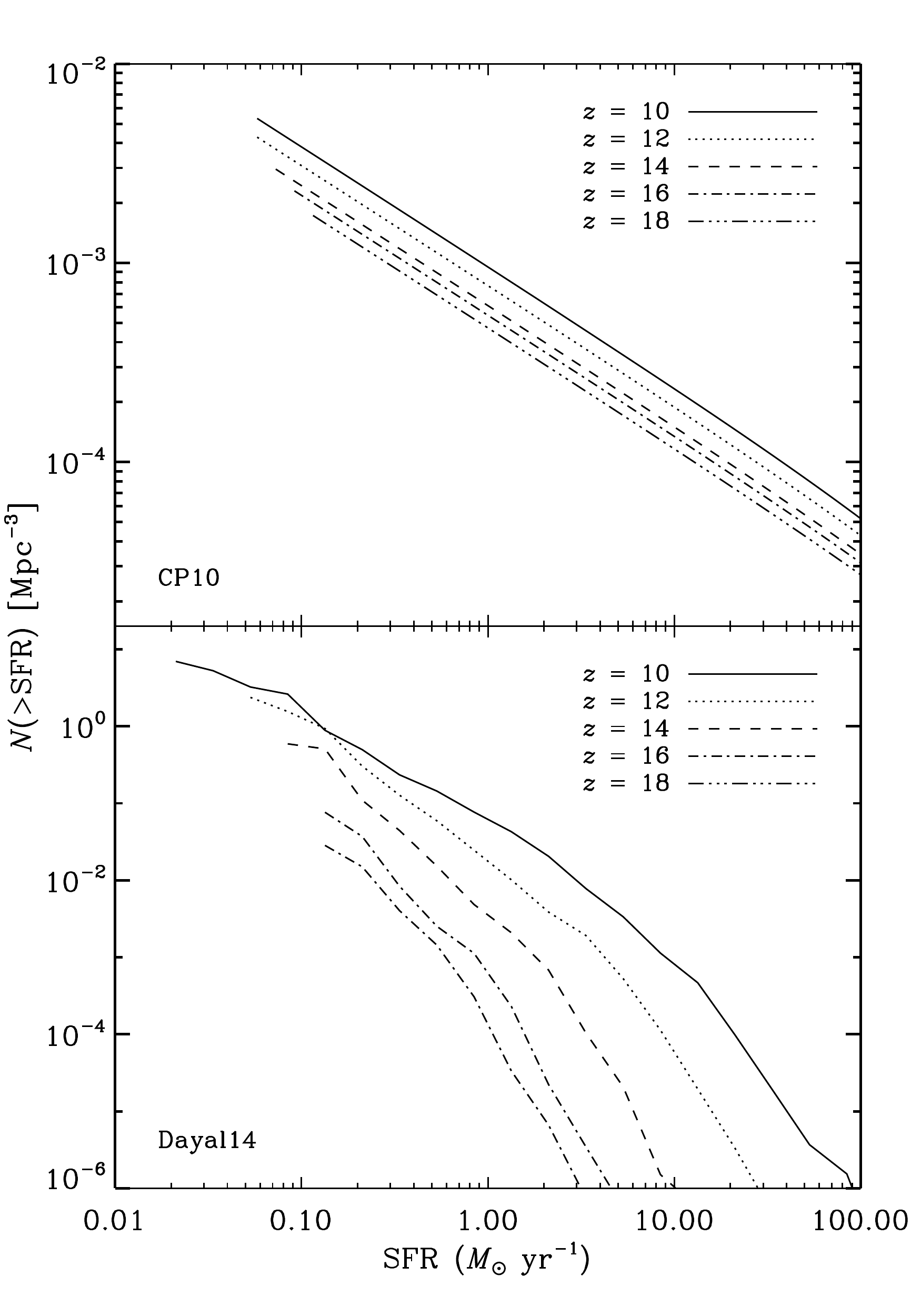}
%\plotone{cum_sfr_z1217.pdf}
\caption{
Comoving number density of galaxies vs. star formation rate and redshift. 
The upper plot is the model of \citet{cp10}.  
The lower plot is the \citet{dayal14} model.  
}
\label{fig:cumNz_cMpc3}
\end{figure}

\begin{figure}[!t]
\centering 
\epsscale{1.2}
\plotone{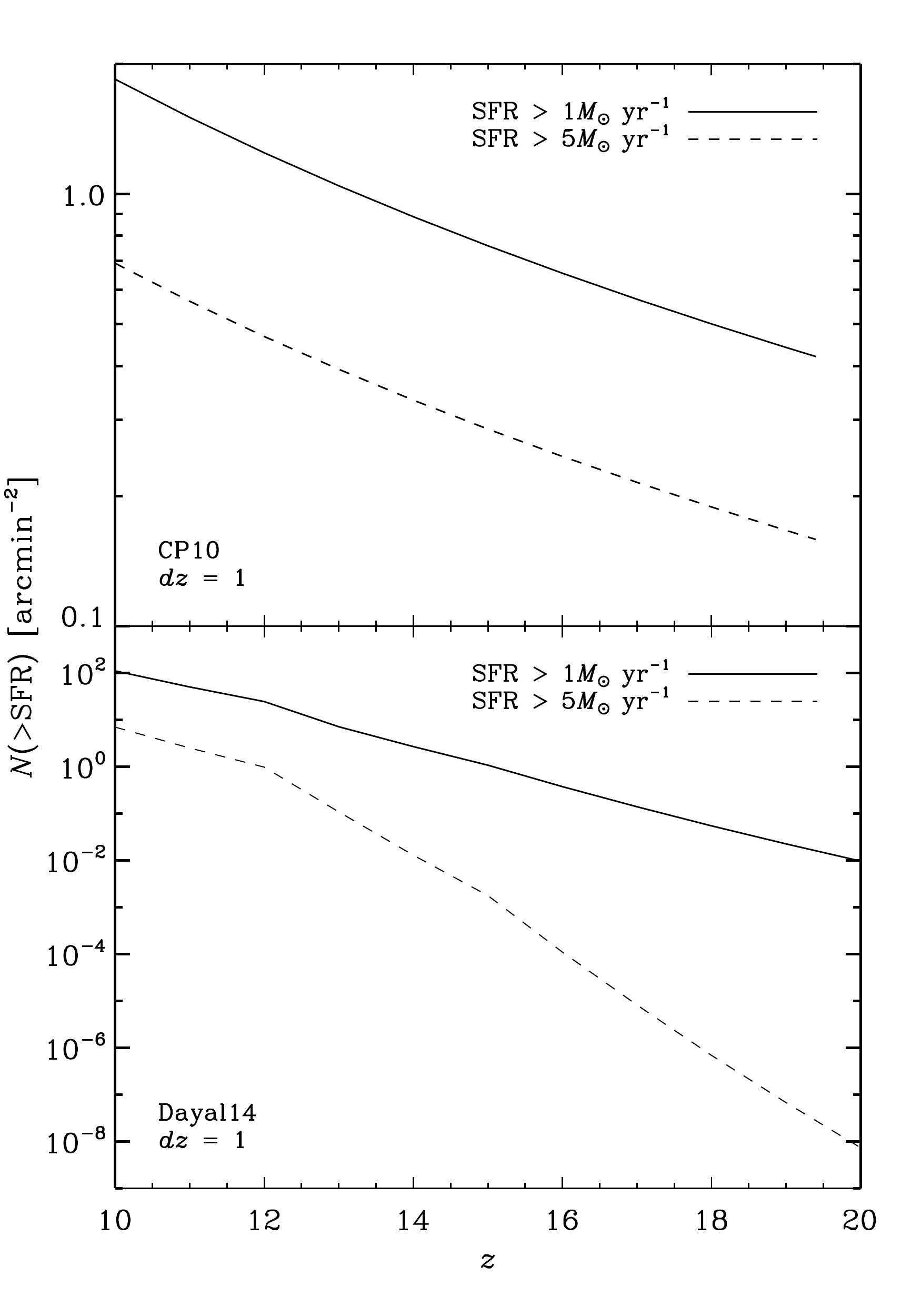}
\caption{
Number of galaxies with star formation rates greater 
than 1\,$M_{\odot}$ yr$^{-1}$ per arcmin$^{2}$ per unit redshift, and 
5\,$M_{\odot}$ yr$^{-1}$ per arcmin$^{2}$ per unit redshift. The upper 
plot is the model of \citet{cp10}. The lower plot is the \citet{dayal14} model.  
}
\label{fig:cumNz}
\end{figure}

Figure \ref{fig:cumNz} shows the number of galaxies per arcmin$^2$ per unit
redshift for SFR $\ge 1\,$$M_\odot$\,yr$^{-1}$ and
5\,$M_\odot$\,yr$^{-1}$, for the CP10 and Dayal14 models,
respectively. The models show markedly different behaviour. The
Dayal14 model has much steeper redshift evolution. The Dayal14 model
also has a much faster drop in density with increasing SFR. Perhaps
fortuitously, at 1\,$M_\odot$\,yr$^{-1}$, the areal densities for the
two models cross at $z \sim 15$.
 
\begin{deluxetable*}{ccccc}
\tablecaption{Number of Detections per 40\,hr Pointing \label{tbl:detrate}
}
\tablewidth{0pt}
\tablehead{
\colhead{Model} & \colhead{ngVLA $z=15$ to 16} & ngVLA $z=15$ to 20 & \colhead{ALMA$^a$ $z=10$ to 10.5} & \colhead{ALMA$^b$ $z = 11$ to 14}
}
\startdata
CP10, 1 M$_\odot$\,yr$^{-1}$ & 0.29 & 1.3 & -- & -- \\
CP10, 5 M$_\odot$\,yr$^{-1}$ & 0.11 & 0.48 & 0.29 & 0.68 \\
Dayal14, 1 M$_\odot$\,yr$^{-1}$ & 0.36 & 0.64 & -- & -- \\
Dayal14, 5 M$_\odot$\,yr$^{-1}$ & $6.9\times 10^{-4}$ & $7.3\times 10^{-4}$ & 2.8 & 1.4 
\enddata
\tablenotetext{a}{Nominal ALMA bandwidth of 8\,GHz}
\tablenotetext{b}{Proposed ALMA bandwidth upgrade to 32\,GHz}
\end{deluxetable*}

The ngVLA can observe the 90\,GHz to 116\,GHz bandwidth
simultaneously, corresponding to $z = 20$ to 15. We also consider just
the number of galaxies between $z = 15$ and 16.  ALMA has receivers
that will cover from $z = 10$ to 15, or frequencies from 173\,GHz to
116\,GHz, but different receivers are needed over the full redshift
range. Currently, the bandwidth is limited to 8\,GHz. We consider an
8\,GHz blind search in the Band 5 from 165\,GHz to 173\,GHz ($z =
10.5$ to 10), and one covering most of Band 4 with a 32\,GHz
bandwidth, from 126\,GHz to 158\,GHz ($z =11$ to 14).

The field of view of the ngVLA at the mean frequency of 100\,GHz is
$\sim 0.38$ arcmin$^{2}$, adopting the FWHM of 0.70arcmin for an 18\,m
antenna.  The field of view of ALMA at the mean frequency of 146\,GHz
is $\sim 0.39$ arcmin$^{2}$, adopting the FWHM of 0.71arcmin for a 12m
antenna.

In Table \ref{tbl:detrate}, we tabulate the number of galaxies
detected in [CII] emission per 40\,hr integration per frequency
tuning, for the ngVLA and ALMA, and for the different models. For the
ngVLA, and for SFR $\ge 1$\,$M_\odot$\,yr$^{-1}$, the models predict
that one to two independent pointings will be required to
detect one galaxy over the full redshift range, on average. For the
CP10 model, these sources have a broader redshift distribution, with
22\% of the sources at $z =15$ to 16. For the Dayal14 model, the
majority (64\%), of the sources are in this lowest redshift bin.

For ALMA and SFR $\ge 5$\,$M_\odot$\,yr$^{-1}$, the predicted number of
detections differs significantly between models. For the 8\,GHz
bandwidth search in Band 5 ($z = 10$ to 10.5), the CP10 model
requires about three pointings for a single detection, on average, while
the Dayal14 model has more low redshift, brighter galaxies, with three sources
per pointing expected. For the 32\,GHz bandwidth search in Band 4 ($z =
11$ to 14), the values are roughly two pointings needed for a single detection
for the CP10 model, and one pointing needed for the Dayal14 model.

Overall, the detection rates in blind surveys will be slow (of order
unity per 40\,hr pointing). However, the observations are well suited
to commensal searches on all programs employing the very wide bands
that may be available in future. Perhaps most importantly, the very
different predictions of the detection rates with respect to redshift
and star formation rate for the two models, both highlights  our lack
of knowledge of the extreme redshift Universe, and implies that the [CII]
results may have great leverage in constraining models of galaxy formation.

As a final note, we point out that blind surveys would be greatly
facilitated by focal plane arrays. This option is being considered for
large interferometric arrays, like the ngVLA, ALMA, and NOEMA,
although it comes at significant expense.  

Large single dish telescopes are also developing large format focal
plane arrays operating in these frequency ranges that will be relevant
for high redshift [CII] searches. The two single dish telescopes that
potentially will have the sensitivity to detect the modest star
formation rate galaxies at $z > 10$ considered herein are the Green
Bank Telescope (GBT\footnote{\url{http://greenbankobservatory.org/}}),
and the Large Millimeter Telescope
(LMT\footnote{\url{http://www.lmtgtm.org/}}).  These telescopes have
comparable sensitivity (within a factor two or better), of ALMA at
100\,GHz.  For example, if a wide-band focal plane array with over 70
elements is deployed at the GBT, the survey speed would then rival,
and possibly surpass, ALMA in the 90\,GHz to 116\,GHz band, depending
on bandwidth.

\subsection{Verifying sources}
\label{sec:verify}

A key issue in blind searches is spurious detections and verifying
sources. For instance, even just considering thermal noise, the
probability for a $\sim 5\sigma$ random positive noise peak is
$2.9\times 10^{-7}$. In the ngVLA blind search at $0.4"$ resolution,
the field of view is $37"$, or a total of 8600 independent synthesized
beams per FoV. The total frequency range covered in 26\,GHz at a
resolution of 100 km s$^{-1}$, or 33\,MHz, hence 780 independent
spectral resolution elements.  The number of independent voxel
elements in the search is then $6.7\times 10^6$, implying two noise
sources by chance. This confusion obviously gets worse at lower
significance, and can be exacerbated by non-Gaussian noise errors due
to sidelobes from continuum or strong line sources \citep{aravena16}.

Recent blind line searches have developed some techniques for making
statistical corrections to number counts based on e.g., comparing the
number of negative and positive detections at a give level
\citep{decarli14, decarli16, walter16, aravena16}. However, the
problem still remains as to how to verify that a given
detection is associated with a $z > 10$ galaxy. For this, other
information will be needed.

One possibility will be broad band near-IR colors from e,g., \textit{JWST}, or
large ground based telescopes. The capability of the \textit{JWST} to make
such measurements has been demonstrated in e.g., \citep{volonteri17}. 
Likewise, follow-up spectroscopy with \textit{JWST} may reveal atomic lines
\citep{barrow17}. Lastly, ALMA could be used to search
for [OIII] emission, in cases of low metallicity galaxies (see below).
  
\subsection{[OIII]\,88\,$\mu$m and [CII]\,158\,$\mu$m }
\label{sec:OIII}

The [OIII] fine structure line at 88\,$\mu$m, which traces ionized
gas, can be as bright as, or brighter than, the [CII] line in low
metalicity galaxies. In their study of nearby galaxies,
\citep{cormier15} show the [OIII]/[CII] ratio has a large scatter, but
the ratio can occasionally be as large as a factor of
$\sim$10. Likewise, ALMA observations of a $z = 7.2$ galaxy show a
similar large [OIII]/[CII] ratio, possibly due to a higher ionization
state for the gas in the galaxy \citep{inoue16}.  On average, the
median [OIII]/[CII] ratio reported for the entire sample of
low-metallicity dwarfs studied by \citep{cormier15} is 2.00 with a
dispersion of 0.36\,dex. We consider a few examples of the comparative
sensitivities herein, adopting this median factor two for low
metallicity dwarfs.

For example, consider the redshift search for a target drop-out
galaxy at $z \sim 12$ with ALMA. The [OIII] and [CII] lines redshift
to frequencies of 260.2 (ALMA Band 6) and 145.8\,GHz (ALMA Band 4),
respectively.  The ALMA sensitivities at these frequencies are
similar, with the rms at 260.2\,GHz being only $\approx$15\% larger
than at 145.8\,GHz (the increase in system temperature with increasing
frequency is offset by the increasing bandwidth for a fixed velocity
width line).  Hence, a factor two stronger [OIII] line requires a
factor $\sim 4$ less observing time to reach a given
signal-to-noise. Countering this factor is the factor 1.8 smaller
fraction bandwidth at the higher [OIII] frequency, thereby requiring 
more frequency tunings in the search.

In terms of blind searches with ALMA itself, again, the factor two
brighter [OIII] line vs. [CII] (in the mean for low metalicity dwarf
galaxies), then requires a factor four less
integration time. However, the primary beam and fractional bandwidth
at 145.8\,GHz are factors of 3.2 and 1.8 larger than at 260.2\,GHz.
Hence, the cosmic volume searched at the higher frequency per tuning
and pointing is a factor 5.7 larger at the lower frequency, 
more than off-setting the signal-to-noise gain, although not by a 
large factor. 

Considering ALMA and [OIII] vs. the ngVLA and [CII] at $z = 15.5$, the
[OIII] and [CII] lines redshift to frequencies of 205 (ALMA Band 5)
and 115\,GHz, respectively.  The sensitivity at 205\,GHz in Band 5 is
a factor of $\approx$11 times worse than at 115\,GHz with the ngVLA.
The primary beam area and fractional bandwidth of the ngVLA at
115\,GHz are factors of 1.4 and 5.8 times larger than at 205\,GHz
with ALMA. The sum total is that ngVLA searches are $\approx$1040 times
faster for ${\rm [OIII]/[CII]} = 1$.  Hence, only if [OIII] is
systematically brighter than [CII] by a factor of $\approx 33$ will ALMA in
[OIII] be competitive in blind searches relative to the ngVLA in
[CII].  It is worth noting that this calculation does not take into
account the fact that ALMA Band 5 observations are severely affected
by the 183\,GHz water line, which essentially wipes out sensitive
observations over the frequency range spanning $\sim 175 -195$\,GHz
($18 \le z \le 16$ for [OIII]).

Overall, for redshift searches on individual candidate drop-out
galaxies, assuming the mean factor two higher signal-to-noise for
[OIII] vs. [CII] for low metallicity dwarf galaxies, the [OIII] line
may prove more effective a tool in terms of search time, although only
marginally since the factor 4 decrease in time due to the stronger
signal is offset by the factor 1.8 smaller fractional bandwidth at
higher frequency. For blind galaxy searches, the increase in
fractional bandwith, and the increase in field of view, at lower
frequency, more than offset the increase in signal to noise, although
not by a large margin.  It is also important to keep in mind that
phase coherence is more of an issue at higher frequency, and hence the
[OIII] line will require better weather.
 
Perhaps most importantly, the various atmospheric windows may
preclude searches at different redshifts for the different
lines. Having two potential lines to search for is an added bonus.

\section{Conclusions}

We have considered observing [CII] 158$\mu$m emission from $z = 10$ to
20 galaxies. The [CII] line may prove to be a powerful tool to determine
spectroscopic redshifts, and galaxy dynamics, for the first galaxies
at the end of the dark ages, such as identified as near-IR dropout
candidates by \textit{JWST}. We emphasize that the nature, and even
existence, of such extreme redshift galaxies, remains at the frontier
of studies of galaxy formation.

In 40\,hr, the ngVLA has the sensitivity to detect the integrated [CII] line
emission from moderate metallicity and (Milky-Way like) star formation
rate galaxies ($Z_{A} = 0.2$, SFR = 1\,$M_\odot$\,yr$^{-1}$), at $z =
15$ at a significance of 6$\sigma$. This significance reduces to
4$\sigma$ at $z= 20$. In 40\,hr, ALMA has the sensitivity to detect the
integrated [CII] line emission from a higher star formation rate
galaxy ($Z_A = 0.2\,Z_{\odot}$, SFR = 5\,$M_\odot$\,yr$^{-1}$), at $z
= 10$ at a significance of 6$\sigma$. This significance reduces to
4$\sigma$ at $z= 15$.  We also consider dependencies on metallically
and star formation rate. Recent studies suggest that the [CII]
luminosity increases rapidly with both metallicity and star formation
rate \citet[see][]{vallini13, vallini15}.  

We perform imaging simulations using a plausible model for the gas
dynamics of disk galaxies, scaled to the sizes and luminosities
expected for these early galaxies. The ngVLA will recover rotation
dynamics for active star-forming galaxies ($\gtrsim
5\,M_\odot$\,yr$^{-1}$ at $z \sim 15$), in reasonable integration
times.

We adopt two models for very high redshift galaxy formation, and
calculate the expected detection rate for [CII] emission at $z \sim
10$ to 20, in blind, wide bandwidth, spectroscopic deep fields. The
detection rates in blind surveys will be slow (of order unity per
40\,hr pointing). However, the observations are well suited to
commensal searches on all programs employing the very wide bands that
may be available in future.  We consider the need for anscillary
observations, such as broad band \textit{JWST} colors or ALMA [OIII] 88$\mu$m
line observations, to verify the association of a given line to a $z >
10$ galaxy.

\vskip 0.2in 

{\sl Acknowledgements}: PD acknowledges support from the European
Research Council's starting grant ERC StG-717001 and from the European
Commission's and University of Groningen's CO-FUND Rosalind Franklin
program. The National Radio Astronomy Observatory is a facilites of
the National Science Foundation operated under cooperative agreement
by Associated Universities, Inc.. We thank Ranga-Ram Chary for discussions
on the models and the paper.

%\clearpage
%\newpage

%\clearpage
%\newpage

\end{document}